\documentclass[12pt,epsfig]{article}

\usepackage{epsf}
\usepackage[dvips]{graphicx}

\begin{document}
\def\bfg #1{{\mbox{\boldmath $#1$}}}
\begin{center}
{\bfseries DYNAMICS OF $^1S_0$ DIPROTON FORMATION IN THE REACTIONS
 $pp\to \{pp\}_s\pi^0$ AND  $pp\to \{pp\}_s\gamma $ IN THE GEV
 REGION }

\vskip 5mm

Yu.N. Uzikov$^{1 \dag}$

\vskip 5mm

{\small
(1) {\it
Joint Institute for Nuclear Researches, LNP, Dubna, Moscow reg. Russian 
Federation
}
\\
$\dag$ {\it
E-mail: uzikov@nusun.jinr.ru
}}
\end{center}

\vskip 5mm
{\small {\bf Abstract}

\vspace{0.3cm}
   COSY-ANKE data on the  cross section of the reactions 
 $pp\to \{pp\}_s\pi^0$ and $pp\to \{pp\}_s\gamma$,
 where $\{pp\}_s$ is the proton pair in the $^1S_0$ state 
 at small excitation energy $E_{pp}<3$ MeV, 
 are analyzed at beam energies $0.5 - 2.0$ GeV within
 the one-pion exchange  model.
 The model includes 
  the subprocesses  $\pi^0 p\to \pi^0 p$ and $\pi^0 p\to \gamma p$
 for the pion- and photo-production, respectively, and accounts for
 the final state pp-interaction.
  A broad maximum  in the energy dependence of
 the $\pi^0$ production  at $\sim 0.5 -0.6$ GeV  and fast increase
 of the $\gamma$-production cross section 
 at $0.3 - 0.55$ GeV observed in the data are
 explained  by the $\Delta$-isobar contribution.   
 An analogy with the dynamics of the deuteron breakup reaction
 $pd\to \{pp\}_sn$ in the $\Delta$-region is outlined. 
}

\section{Introduction}

Quasi-binary reactions $AB\to \{pp\}_sC$ with formation of a proton pair
at small excitation energy $E_{pp}=0-3$ MeV, i.e. the $^1S_0$
diproton $\{pp\}_s$, 
are of great interest  at high transferred momenta
since transition amplitudes of these reactions
require high momentum components of the pp-wave function. In comparison 
to very similar (in kinematics) reactions $AB\to dC$ with the final 
deuteron $d$,
the reactions with the diproton are expected to give more definite
 information on short-range NN-dynamics. The reason is that 
 the contribution of non-short range mechanisms related to excitation of 
 the $\Delta$-isobars in intermediate states 
 is expected to be strongly 
 suppressed for the $AB\to \{pp\}_sC$ reactions as compared to the 
$AB\to dC$
 due to isospin symmetry and conservation of angular momentum and parity. 
 So, in the reaction $pd\to \{pp\}_sn$ this suppression
 is given by the factor $\frac{1}{9}$ \cite{uz2002}.
 Furthermore,
 in the reaction  $pp\to \{pp\}_s\pi^0$ the intermediate
 S-wave $\Delta N$ state is completely forbidden \cite{niskanen}.
 Similarly, in the $pp\to \{pp\}_s\gamma$ reaction 
 direct excitation of the $\Delta-$isobar,
 dominating the $\gamma d\to pn$ reaction via $M1$ transition
is also forbidden.
  
 Contrary to those expectations, the cross section of the reactions 
 $pp\to \{pp\}_s\pi^0$ \cite{kurbatov} and  $pp\to \{pp\}_s\gamma $
 \cite{komar08}
 recently measured in forward direction
 for beam energies  $0.5 - 2.0$ GeV and $0.35 - 0.55$ GeV, respectively,
  demonstrate prominent peaks in the $\Delta(1232)$-isobar region.
 In the deuteron breakup reaction $pd\to \{pp\}_sn$ measured in
 Ref.\cite{komar2003} the $\Delta(1232)$ peak is non-visible
 in the energy dependence of the cross section for the
 backward scattered neutron, 
 however,  theoretical analyses \cite{jhuz2003,uzjhcw} suggest,
 that the $\Delta$
 contribution dominates in this reaction at $0.5 - 1.3$ GeV.

 Observation of the $\Delta$ peaks  in the data on the  reactions
 $pp\to \{pp\}_s\pi^0$ \cite{kurbatov} and  $pp\to \{pp\}_s\gamma $
 \cite{komar08}
 would mean that the high momentum component of the NN-wave
 function, which might be
 hidden by the $\Delta$- contribution in the corresponding
 reactions with the deuteron,
is actually rather week. In other words,  new data \cite{kurbatov,komar08},
 most likely,
confirm the result of the previous analysis of the reaction $pd\to\{pp\}_sn$
\cite{jhuz2003}, which suggests  softness of the NN-interaction potential
 at short distances.
 To study this conjecture theoretical analysis is required.

 Here we present the results of 
  calculations of the differential cross sections of the reactions
 $pp\to \{pp\}_s\pi^0$ and $pp\to \{pp\}_s\gamma$
 at $\theta_{cm}=0^\circ$ 
 within the one-pion exchange model, which includes
 the subprocesses $\pi^0 p\to \pi^0 p$ and $\pi^0 p\to \gamma p$, 
 respectively. A similar model with the subprocess
 $\pi^0 d\to np$ was applied  earlier to the reaction $pd\to \{pp\}_sn$ 
 \cite{uzjhcw}.

\section{The deuteron breakup reaction  $pd\to \{pp\}_sn$}

The deuteron breakup reaction $pd\to \{pp\}_sn$
 was studied  at COSY 
\cite{komar2003}.  A theoretical analisys \cite{jhuz2003} was performed 
 within the sum of the
 following mechanisms:
 one-nucleon-exchange (ONE) with initial and final state interaction  included,
 $\Delta$-isobar excitation ($\Delta$) and
 single-scattering (SS). 
%
 \begin{figure}[hbt]
\begin{minipage}[c]{100mm}
\includegraphics[width=100mm]{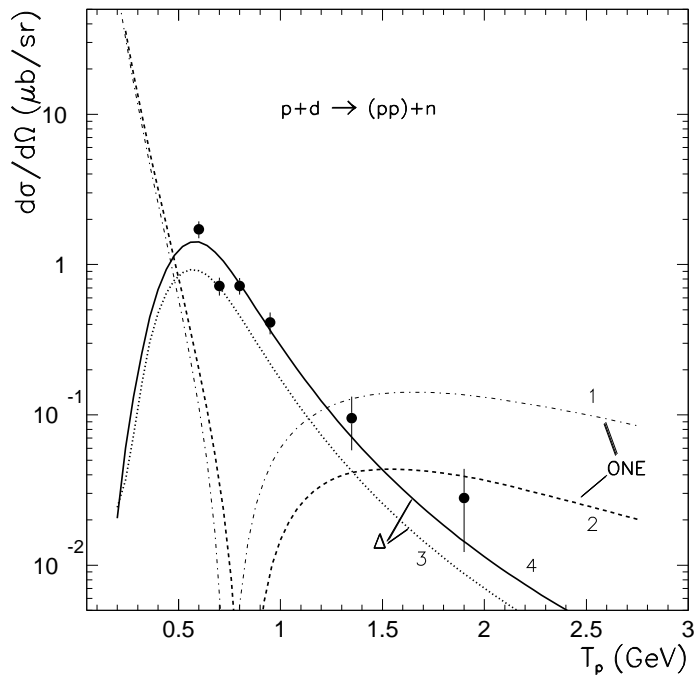}
\end{minipage}
\hspace{\fill}
\caption{ The $pd\to \{pp\}_sn$ data \cite{komar2003} ($\bullet$)
 at $\theta_{cm}^n=180^\circ$
 verus the beam energy in comparison with  the  ONE (1, 2)  and $\Delta$
 mechanisms (3,4) calculations from Ref. \cite{jhuz2003}.
 Replacement of the Paris NN potential (1,3) by the
 CD Bonn one (2,4), decreases the ONE contribution, but increases
 the $\Delta$-contribution.
}
\label{figpd}
\end{figure}
 The analysis shows (see Fig. \ref{figpd})
 that  at 0.8 GeV the ONE mechanism
 has a minimum due to repulsive core in the NN-interaction at $r_{NN}\sim 0.5
 $fm, but
 the $\Delta$ contribution has a maximim  at 0.6 GeV and completely
 dominates this reaction.
 This $\Delta$-maximum is
  not visible  as a bump in the cross section due to a  large ONE
 contribution below the ONE-node. However, only rather soft
 NN-interaction potential  like the CD Bonn one \cite{Machleidt}
 provides agreement with the data.
 When replacing  a hard NN-interaction potential
(RSC \cite{Reid}, Paris \cite{paris}) by the soft one
(CD Bonn) 
 the ONE contribution  decreases, whereas
 the $\Delta$ contribution increases providing agreement with the
 COSY data.
 On the other hand,
 more hard NN-models like Paris and especially RSC provide too
 large magnitude of the  high momentum components of the NN-wave function
 and, therefore,  lead to
 strong contradiction with the data especially above 1 GeV
 (see Ref. \cite{jhuz2003}).
 Further analysis \cite{uzjhcw}
  within the OPE model with the subprocess
 $\pi ^0d\to pn$ provided an independent confirmation
 of the dominant contribution
 of the $\Delta(1232)$-isobar in this reaction at $0.5 - 1$ GeV and suggested
 sizable  admixture of the ONE mechanism compatible with the CD Bonn model.

\section{The  reaction  $pp\to \{pp\}_s\pi^0$}

 The reaction
  $pp\to \{pp\}_s\pi^0$ is the
 simplest inelastic process in the pp-collision, which can reveal
 underlying dynamics of NN interaction.
 Restriction to only one pp-partial wave
 (s-wave) in the final state
 considerably simplifies a comparison with theory.
 The reaction $pp\to \{pp\}_s\pi^0$
 is very similar kinematically to the reaction $pp\to d\pi^+$, but its
 dynamics can be essentially different. In fact, quantum numbers of the
 diproton state ($J^\pi=0^+,\,I=1,\, S=0, \, L=0$) differ from these for
 the deuteron ($J^\pi=0^+, I=0,\, S=1, L=0,2$). Therefore, transition
 matrix elements for these two reactions are also different.
  Due to the generalized Pauli principle and
   angular momentum and P-pariry conservation
 only negative parity states  are allowed in the reaction
 $pp\to \{pp\}_s\pi^0$. Thus, for the intermediate $\Delta N$ state 
 only odd partial 
waves 
 are allowed.
 In contrast, in the 
$pp\to d\pi^+$ reaction both negative and positive parity  states 
are allowed and formation of the intermediate S-wave $\Delta N$ state
 with $J^P=2^+$ leads to a  perfect resonance looping in the $^1D_2$ 
$pp$-partial wave in the respective Argand diagram \cite{arndt}.
 Therefore, the relative contribution of the 
$\Delta$-mechanism to the reaction $pp\to \{pp\}_s\pi^0$ is expected to be
suppressed as compared to the reaction $pp\to d\pi^+$. This argument was
 applied in Ref. \cite{uzwilk2001} to explain   a very 
 small ratio 
(less of few percents) of the spin-singlet to spin-triplet
 pn-pairs observed in the LAMPF data \cite{HGabitch}
 in the final state interaction region of the reaction
 $pp\to pn\pi^+$ at proton beam energy 0.8 GeV. 
 Furthermore, since $\Delta-$type 
mechanisms are of long-range type,
 reduction of their contribution would mean that other mechanisms, like
$N^*$-exchanges \cite{sharmamitra} which are more sensitive to short-range
 NN-dynamics, could be more pronounced in the reaction
 $pp\to \{pp\}_s\pi^0$ as compared to 
the $pp\to d\pi^+$ reaction \cite{ponting}.
\begin{figure}[h]
 \epsfysize=40mm
 \centerline{
 \epsfbox{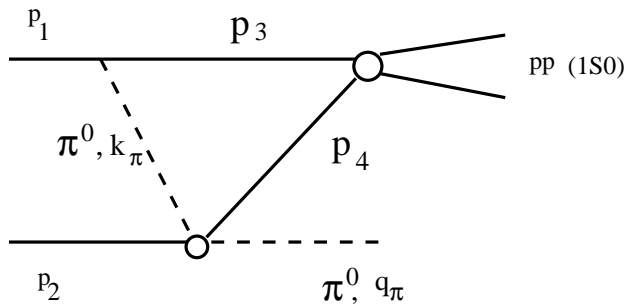}}
 \caption{The OPE mechanism of the rection $pp\to  \{pp\}_s\pi^0$.}
\label{fig1}
\end{figure}

  The cross section of the reaction $pp\to \{pp\}_s\pi^0$ was measured
  recently
  at energy  0.8 GeV in Ref.\cite{dymov06} and  
  at beam energies $0.5 - 2.0$ GeV in Ref. \cite{kurbatov}.
  At  zero angle, the data \cite{kurbatov} show 
  a broad maximum in the energy dependence of
  the cross section at $0.5 - 1.4$ GeV. This maximum is similar in shape and 
  position to the well known $\Delta-$ maximum in the reaction
 $pp\to d\pi^+$.  
  However, a comparison with the  microscopical model  calculation 
 \cite{niskanen}, which  includes
 $\Delta(1232)$-isobar  excitation and s-wave $\pi N$-rescattering,
 reveals  very strong disagreement between the model and the data
 \cite{kurbatov}  at
 energies $0.5 - 0.9$ GeV  both in the absolute value
 and shape of energy dependence  of the cross section.

\begin{figure}[hbt]
\begin{minipage}[c]{100mm}
\includegraphics[width=100mm]{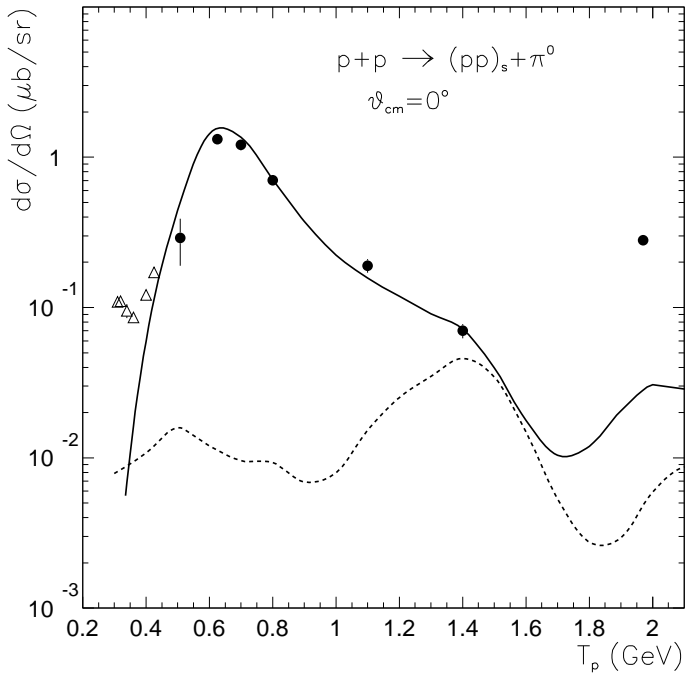}
\end{minipage}
\hspace{\fill}
\caption
 {The  differential cross section of the reaction
  $pp\to \{pp\}_s\pi^0$ versus the beam energy   at $\theta_{cm}=0^\circ$.
 The OPE model (full line)
 is compared with the data:  $\bullet$ -- \cite{kurbatov}, triangles --
\cite{bilger}.
 The dashed curve shows the OPE result obtained without the isospin $3/2$
 contribution to the amplitude $\pi^0 p\to \pi^0 p$. The calculated cross
  sections are scaled by the factor 1/6 (see text).
 }
\label{figpi0}
\end{figure}

  Here \cite{uzpi02008} we analyse these data employing a simpler model,
  which includes the subprocess $\pi ^0 p\to \pi^0 p$ and the final
  state pp$(^1S_0)$-interaction (Fig.\ref{fig1}). The formalism is very
  similar to that developed for the $pd\to \{pp\}_sn$ reaction \cite{uzjhcw}.
  We use the impulse approximation, i.e.
 the  amplitude of the reaction $\pi^0p \to \pi^0p$ is taken
 off the loop inegral sing.
 Therefore, the cross section
  of the  reaction $pp\to \{pp\}_s\pi^0$  in forward direction is proportional
  to the forward cross section of the reaction  $\pi^0p \to \pi^0p$
  taken from the data \cite{arndt}. The structure formfactor is calculated 
  using the CD Bonn model for pp-interaction \cite{Machleidt}.
 The cutoff parameter for the
  monopole formfactor in the $\pi NN$ vertex is taken as $\Lambda=0.65$ GeV/c.
  The results presented  in Fig.\ref{figpi0} by full line show
  that the observed shape of the peak
  is in agreement with the dominance of
  the $\Delta(1232)$-isobar contribution. Indeed, 
  exclusion of the the isospin $3/2$
  contribution  from the  amplitude of  reaction $\pi^0p \to \pi^0p$ 
  (dashed curve)  leads to  strong disagreement with the data.
  In absolute value the OPE cross section overestimates the data by factor
  of 6. The main part of this factor  can be explained by the employed
 impulse approximation. Indeed, within the impulse approximation
  one cannot exclude  intermediate $\Delta N$-states of positive parity,
  which  are forbidden in this reaction. In order to exclude these states one
  needs to consider the 
  $\Delta$-isobar ecxitation explicitely.

\section{The  reaction  $pp\to \{pp\}_s\gamma$}

\begin{figure}[hbt]
\begin{minipage}[c]{90mm}
\includegraphics[width=90mm]{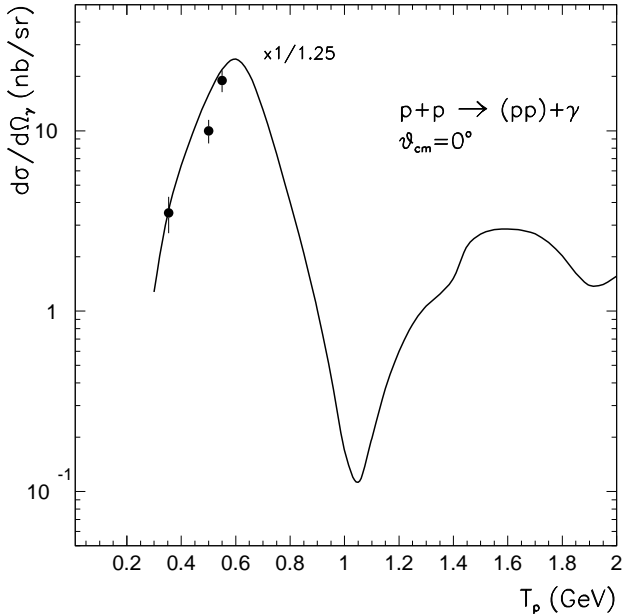}
\end{minipage}
\caption{The forward differential cross section
 $pp\to \{pp\}_s\gamma$ in comparison with the OPE model (curve
 scaled by the factor 0.8). 
  Data ($\bullet$) are taken from  Ref. \cite{komar08}.
}
\label{figgamma}
\end{figure}

 Another simplest process which allows to probe fundamental
 properties of NN system is photoabsorption on two nucleon systems.
  The deuteron photodisintegration reaction $\gamma d\to pn$ is widely used
 as a testing ground for different theoretical models of the NN-interaction,
 however, much less is known on the photodisintegration of the  diproton,
 $\gamma \{pp\}_s\to pp$, or the inverse process of the photoproduction
 $pp\to  \{pp\}_s\gamma$. Whereas in the photodisinegration of the deuteron
 the M1 magnetic  dipole transition dominates at several hundred MeV  through
 the excitation of the $\Delta(1232)$ isobar, in the reaction with the
 $^1S_0$ diproton M-odd multipoles are forbidden due to angular momentum and
 parity conservation. Therefore, there is no direct
  contribution of the intermediate S-wave
 $\Delta N$ states
 in the reaction $pp\to \gamma \{pp\}_s$.  Non-direct excitation of the
 $^5S_2$ $\Delta N$ state is possible via the E2 transition \cite{WNA},
 but  this contribution is expected to be  less
 important
 than the M1-transition. 
   The OPE model of the reaction $pp\to  \{pp\}_s\gamma$ allows to
 account for the $\Delta$ contributions via the subprocess 
$\pi^0 p\to p\gamma$. The corresponding OPE diagram is
 similar to those in Fig. \ref{fig1}, but with the
 subproscess $\pi^0 p\to p\gamma$ in the down vertex.
 The result of the OPE calculations are shown in Fig.\ref{figgamma}.
 One can see that this model explaines the observed in Ref. \cite{komar08}
 rise of the cross section almost quantatively. The  second bump  at
 1.6 GeV  is caused by the energy dependence of the $\pi^0 p\to p\gamma$
 cross section \cite{arndt} and related to excitation of more heavy nucleon
 isobars.

\section{Conclusion}

   Parity and angular momentum conservation exclude
   the S-wave $\Delta N$-intermediate state 
  from the reaction $pp\to \{pp\}_s\pi^0$.
  In  similar way, the M1  transition, dominating in the
  $\gamma d\to pn$  reaction at several hundred MeV via excitation of the
  $\Delta$-isobar, is forbidden in the reaction
  $pp\to \{pp\}_s\gamma$. This suppression is similar to that in the reaction
 $pd\to \{pp\}_sn$ \cite{uz2002,jhuz2003} as compared to the $pd\to dp$ process.
 Therefore,
 one could expect that some
  features of  the short-range dynamics, which, perhaps, are not
  visible in the reactions with
  deuteron, $pp\to d\pi^+$ and $\gamma d \to pn$,  may reveal themselves in the
  corresponding reactions with the diproton. The OPE calculations, in
  agreement with the data show,  however, that the $\Delta$-contribuition
  is still significant in the reactions
 $pp\to \{pp\}_s\pi^0$ and
 $pp\to \{pp\}_s\gamma$. It would  mean that short-range
  effect is rather minor itself  in these reactions
  in the considered region.

\end{document}